\documentclass[a4paper,aps,twocolumn,prl,showpacs, amsmath,amssymb,superscriptaddress, 12p6]{revtex4-1}

\usepackage{graphicx}
\usepackage{dcolumn}
\usepackage{bm}
\usepackage{longtable}
\usepackage{color}

\makeatletter
\newcommand\figcaption{\def\@captype{figure}\caption}
\newcommand\tabcaption{\def\@captype{table}\caption}

\makeatother

\begin{document}




\title{Dimensionality-driven metal-insulator-transition in spin-orbit coupled SrIrO$_3$ }

\author{P. Sch{\"u}tz}
\affiliation{Physikalisches Institut and R\"ontgen Center for Complex Material Systems (RCCM), Universit\"at W\"urzburg, Am Hubland,
D-97074 W\"urzburg, Germany}
\author{D. Di Sante}
\affiliation{Institut f\"ur Theoretische Physik und Astrophysik, Universit\"at W\"urzburg, Am Hubland,
D-97074 W\"urzburg, Germany}
\author{L. Dudy}
\author{J. Gabel}
\author{M. St{\"u}binger}
\author{M. Kamp}
\affiliation{Physikalisches Institut and R\"ontgen Center for Complex Material Systems (RCCM), Universit\"at W\"urzburg, Am Hubland,
D-97074 W\"urzburg, Germany}
\author{Y. Huang}
\affiliation{Van der Waals - Zeeman Insitute, University of Amsterdam, Science Park 904, 1098 XH Amsterdam, The Netherlands}
\author{M. Capone}
\affiliation{CNR-IOM-Democritos National Simulation Centre and International School for Advanced Studies (SISSA), Via Bonomea 265, I-34136 Trieste, Italy}
\author{M.-A. Husanu}
\affiliation{National Institute of Materials Physics,  Atomistilor 405 A, 077125 Magurele, Romania}
\affiliation{Swiss Light Source, Paul Scherrer Institut, CH-5232 Villigen, Switzerland}
\author{V. Strocov}
\affiliation{Swiss Light Source, Paul Scherrer Institut, CH-5232 Villigen, Switzerland}
\author{G. Sangiovanni}
\affiliation{Institut f\"ur Theoretische Physik und Astrophysik, Universit\"at W\"urzburg, Am Hubland,
D-97074 W\"urzburg, Germany}
\author{M. Sing}
\author{R. Claessen}
\affiliation{Physikalisches Institut and R\"ontgen Center for Complex Material Systems (RCCM), Universit\"at W\"urzburg, Am Hubland,
D-97074 W\"urzburg, Germany}

\date{\today}

\begin{abstract}%
Upon reduction of the film thickness we observe a metal-insulator transition in epitaxially stabilized, spin-orbit coupled SrIrO$_3$ ultrathin films.  By comparison of the experimental electronic dispersions with density functional theory at various levels of complexity we identify the leading microscopic mechanisms, i.e., a dimensionality-induced re-adjustment of octahedral rotations, magnetism, and electronic correlations. The astonishing resemblance of the band structure in the two-dimensional limit to that of bulk Sr$_2$IrO$_4$ opens new avenues to unconventional superconductivity by "clean" electron doping through electric field gating.
\end{abstract}

\maketitle
\begin{figure}%
\centering%
\includegraphics[width = \linewidth]{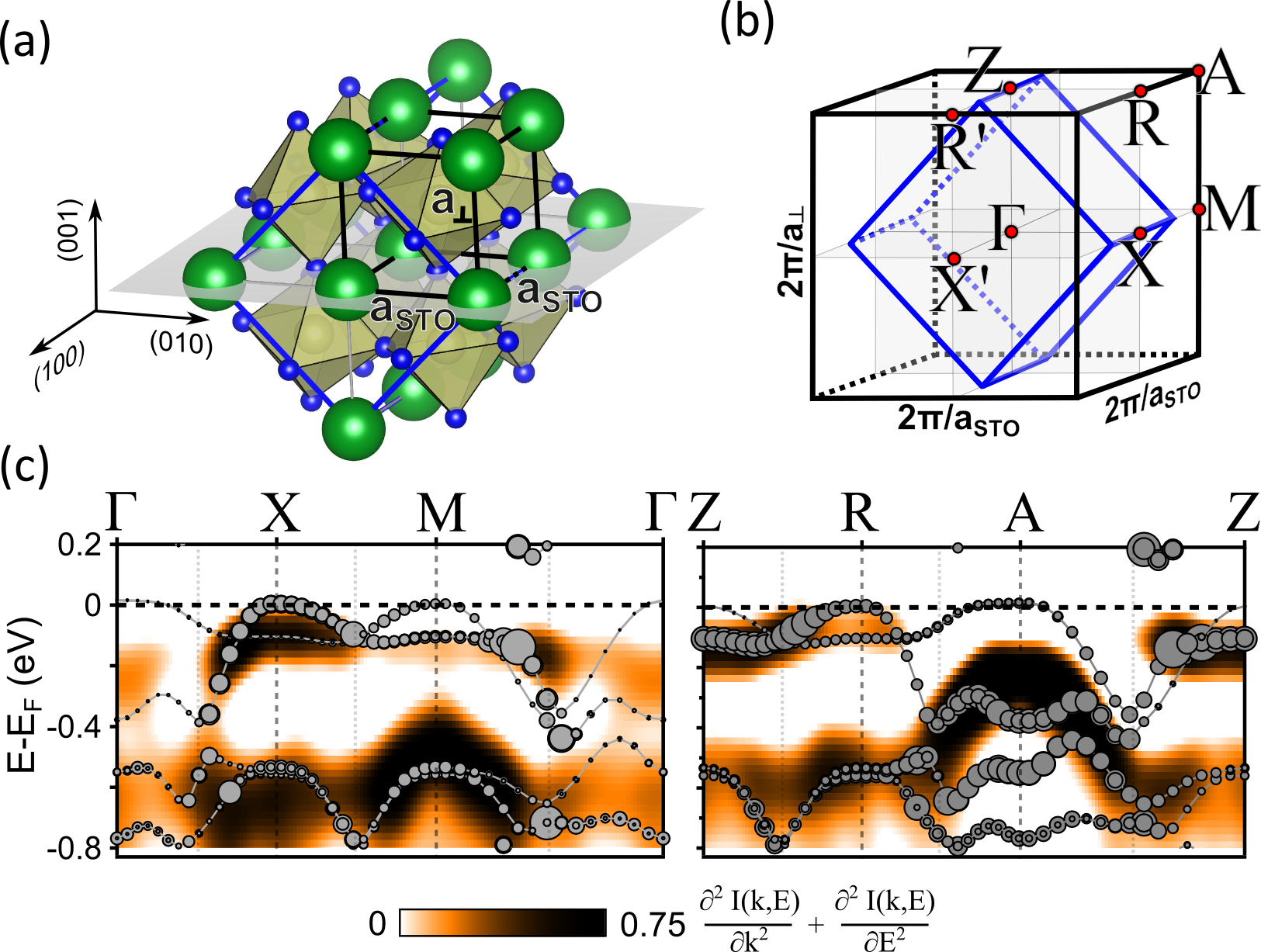}%
\caption{(Color online) (a) Real-space lattice structure of SrIrO$_3$ including octahedral rotations ($a^+b^-b^-$  in Glazer notation with the $a$-axis orthogonal to the film surface normal) and strain. The orthorhombic unit cell (blue) is enlarged by $2\times\sqrt{2}\times\sqrt{2}$ with respect to the tetragonal unit cell (black). (b) Reciprocal space structure of the orthorhombic (blue) and tetragonal (black) structure. (c) SX-ARPES bandmaps along the pseudo-tetragonal high-symmetry lines $\Gamma$-X-M-$\Gamma$ and Z-R-A-Z in comparison to DFT+$U$ band structure calculated in the orthorhombic setting and unfolded into the tetragonal Brillouin zone. }
\label{fig:OctRot}%
\end{figure}
\begin{figure*}%
\centering%
\includegraphics[width =  \linewidth]{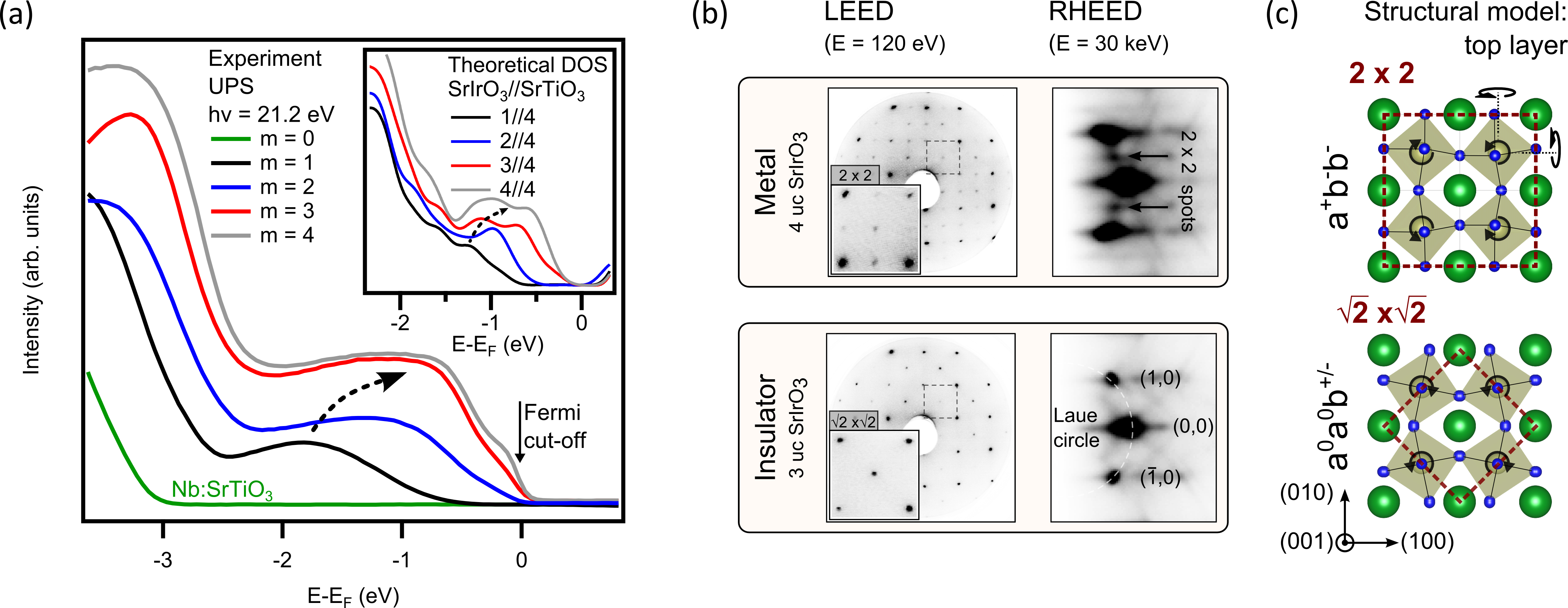}%
\caption{(Color online) (a) Ultraviolet photoelectron spectroscopy (UPS) of SrIrO$_3$ films with thickness $m$ (bare Nb:SrTiO$_3$, $m = 1,2,3,4~\text{uc}$) exhibiting the opening of a charge gap between 3 and 4 uc. Inset: DOS from DFT+$U$ slab calculations of $m$ SrIrO$_3$ layers on 4 SrTiO$_3$ layers ($m$//4). (b) LEED and RHEED patterns of an insulating ($m=3$) and a metallic ($m=4$) SrIrO$_3$ film exhibit a structural transition from a $\sqrt{2}\times\sqrt{2}$ to a $2\times2$ surface periodicity across the metal-insulator-transition. (c) Structural model of the film surface, explaining the $\sqrt{2}\times\sqrt{2}$/ $2\times2$ surface periodicity as result of $a^0a^0b^{+/-}$/ $a^+b^-b^-$ IrO$_6$ octahedral rotations in the film.} %
\label{fig:MIT}%
\end{figure*}
\begin{figure*}%
\centering%
\includegraphics[width =  0.8\linewidth]{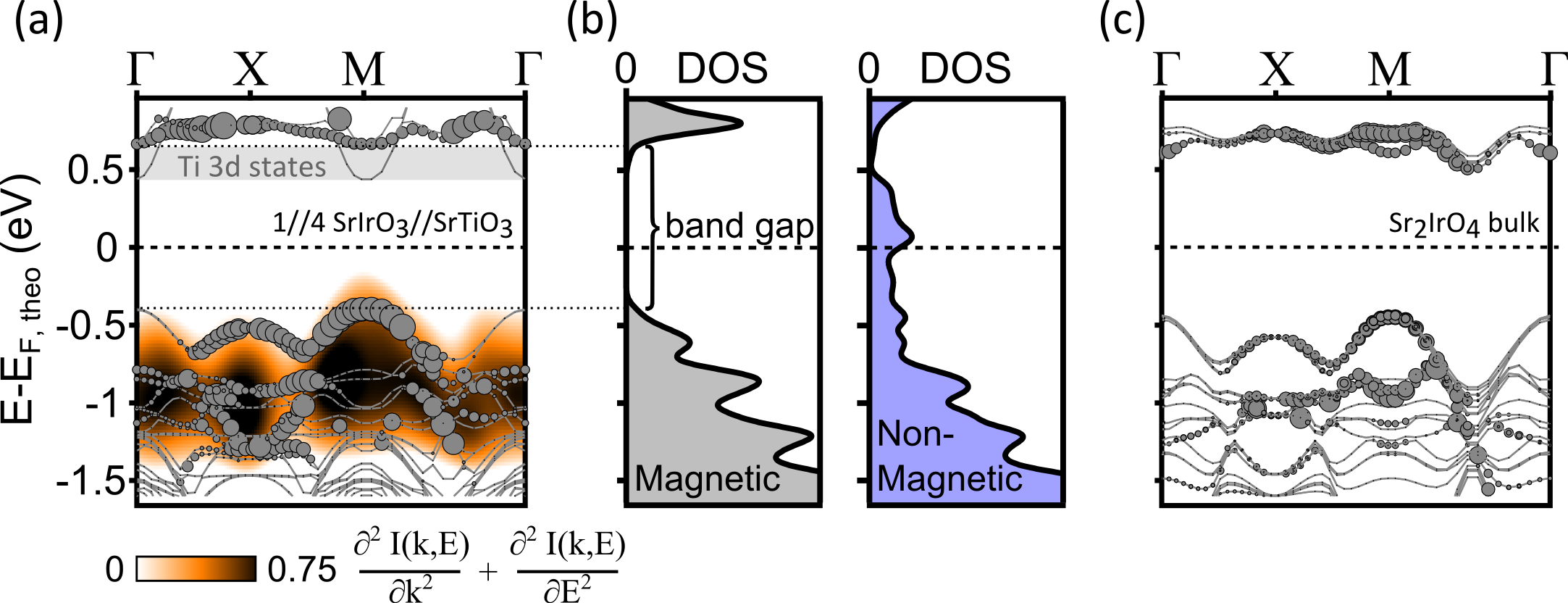}%
\caption{(Color online) (a) SX-ARPES bandmap of a SrIrO$_3$ monolayer grown on Nb:SrTiO$_3$ and DFT+$U$ bandstructure of a $1//4$ SrIrO$_3$//SrTiO$_3$ slab along the pseudo-tetragonal high-symmetry line $\Gamma$-X-M-$\Gamma$. (b) DFT+$U$ $k$-integrated density of states (DOS) of the $1//4$ slab. The antiferromagnetic solution exhibits a charge gap of $\approx 1\,\text{eV}$, while the non-magnetic solution is in a metallic state. (c) DFT+$U$ band structure calculation for bulk Sr$_2$IrO$_4$.}%
\label{fig:Insulator}%
\end{figure*}

\noindent Although typically viewed as disparate properties, the interplay between strong spin-orbit coupling (SOC) and electronic correlations in high-$Z$ $5d$ transition metal oxides can lead to exotic quantum states of matter like Kitaev spin liquids \cite{jackeli_mott_2009, witczak-krempa_correlated_2014} and topological phases \cite{pesin_mott_2010,zhang_actinide_2012}. Sr$_2$IrO$_4$, the quasi-two-dimensional member ($n=1$) of the layered-perovskite Ruddlesden-Popper series Sr$_{n+1}$Ir$_n$O$_{3n+1}$, has attracted special attention as a possible parent material for exotic superconductivity since it exhibits a SOC-driven Mott insulating state \cite{kim_novel_2008}, which largely reproduces the fermiology of hole-doped cuprates upon electron-doping \cite{kim_fermi_2014, kim_observation_2016}. In contrast, its metastable three-dimensional ($n=\infty$) counterpart SrIrO$_3$ exhibits a semimetallic state \cite{nie_interplay_2015} believed to be in proximity to a dimensionality-driven metal-insulator transition (MIT) \cite{moon_dimensionality-controlled_2008}.\\
In this Letter, we investigate the electronic and structural properties of epitaxially-grown ultrathin perovskite SrIrO$_3$ films by tuning the film thickness with atomic precision. We observe the opening of a distinct charge gap at the chemical potential and concurrent changes in the film crystalline structure when approaching the two-dimensional limit. In a combined experimental and theoretical approach using soft x-ray angle-resolved photoelectron spectroscopy (SX-ARPES) and \textit{ab initio} density functional theory (DFT and DFT$+U$) calculations we investigate the evolution of the electronic band structure across the film thickness-driven MIT and shed light on the complex interplay between electronic correlations, structural degrees of freedom, magnetism, and dimensionality.\\%
%
SrIrO$_3$ thin films were heteroepitaxially grown on TiO$_2$-terminated SrTiO$_3$ (001) substrates by pulsed laser deposition (PLD) from a polycrystalline SrIrO$_3$ target. The films adopt a pseudo-tetragonal perovskite structure with an in-plane lattice constant locked to that of SrTiO$_3$ ($a=3.905\,\text{\AA}$) and an out-of-plane lattice constant of $3.99\,\text{\AA}$. Due to collective rotations of the IrO$_6$ octahedra ($a^+b^-b^-$ in Glazer notation with the $a$-axis along the [100] or [010] direction of the substrate \cite{longo_structure_1971,nie_interplay_2015}) the real-space unit cell is enlarged by $2\times\sqrt{2}\times \sqrt{2}$ with respect to the tetragonal unit cell (for a thorough structural characterization, see Supplemental Material \cite{Supplemental}). SX-ARPES measurements were performed at $20\,\text{K}$ at the ADRESS beamline of the Swiss Light Source, Paul Scherrer Institute \cite{strocov_high-resolution_2010,strocov_soft-x-ray_2013}. Density functional theory calculations were performed by using the VASP \textit{ab initio} simulation package \cite{kresse_efficient_1996} within the projector-augmented-plane-wave (PAW) method \cite{blochl_projector_1994, kresse_ultrasoft_1999}, using the generalized gradient approximation (GGA) as parametrized by the PBE GGA functional \cite{perdew_generalized_1996}. Spin-orbit coupling was self-consistently included \cite{steiner_calculation_2016} and the Coulomb repulsion $U$ and exchange interaction $J$ of Ir~$d$ orbitals were treated within the rotationally invariant DFT+$U$ scheme of Liechtenstein \textit{et al.} \cite{liechtenstein_density-functional_1995}. \\
\begin{figure*}%
\centering%
\includegraphics[width =  \linewidth]{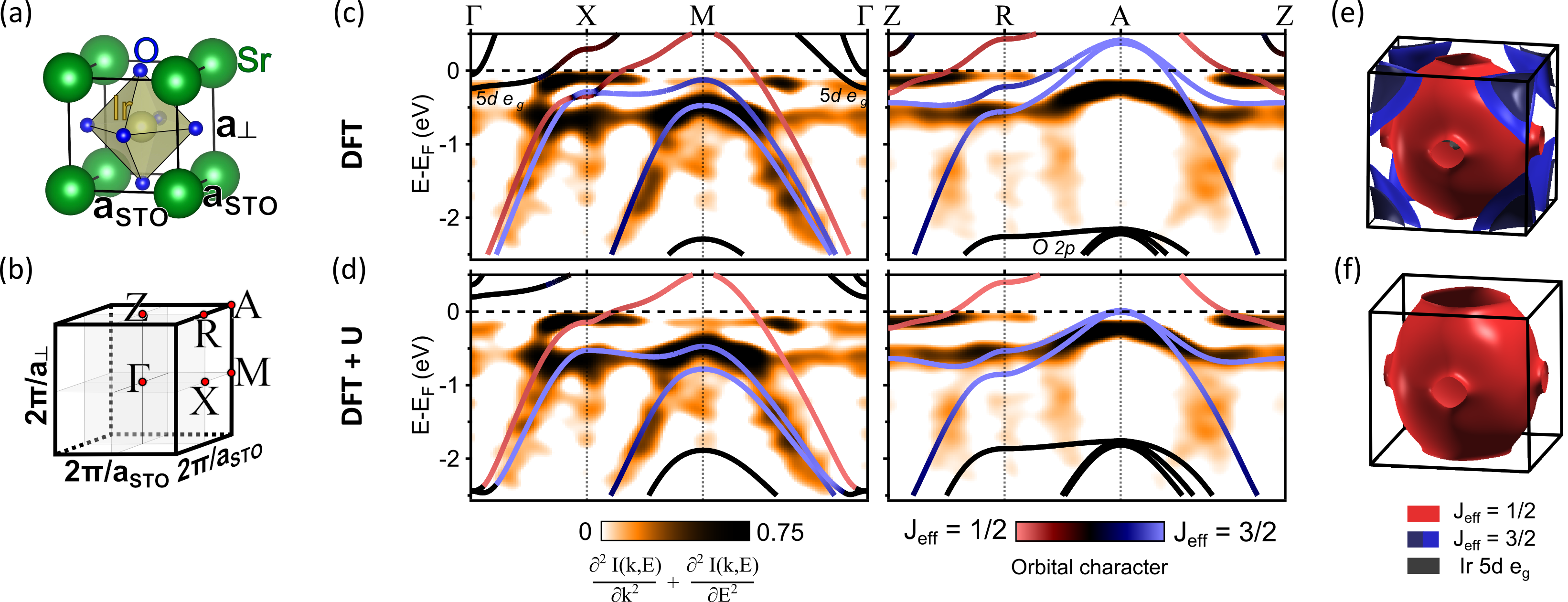}%
\caption{(Color online) (a) Real and (b) reciprocal space structure of strained, tetragonal SrIrO$_3$ without octahedral rotations. (c), (d) $E$ vs $k$ dispersions along the high-symmetry lines $\Gamma$-X-M-$\Gamma$ and Z-R-A-Z measured by SX-ARPES ($h\nu = 745\,\text{eV}$ and $h\nu= 660\,\text{eV}$, respectively) and compared to DFT+$U$ calculations. The band structure was calculated for the tetragonal setting and projected onto a $J_\text{eff}= (1/2, 3/2)$ basis with (c) $U=0\,\text{eV}$ and $J=0\,\text{eV}$ and (d) $U=3.4\,\text{eV}$ and $J=0.4\,\text{eV}$. The introduction of a sizeable on-site Coulomb repulsion significantly enhances the agreement between theory and experimental results. (e), (f) Fermi surface topology without (e) and with (f) on-site Coulomb repulsion $U$ and exchange coupling $J$.}
\label{fig:arpes}%
\end{figure*}%
%
Figure~\ref{fig:arpes} shows the experimental and theoretical band structures along the high-symmetry lines $\Gamma$-X-M-$\Gamma$ and Z-R-A-Z obtained from SX-ARPES on a 9 unit cells (uc) thick metallic SrIrO$_3$ film and paramagnetic DFT($+U$) calculations for the bulk material. As starting point for the analysis of the photoemission data we consider a simplified tetragonal perovskite structure compressively strained to the SrTiO$_3$ substrate ($a_\perp/a_{\text{STO}} \approx 1.02$) as depicted in Fig.~\ref{fig:arpes}\,(a) and (b). We investigate the combined effect of on-site Coulomb repulsion $U$ and exchange-coupling $J$ on the DFT($+U$) band structure and Fermi surface topology. The resulting bands are projected onto a $J_\text{eff}=(1/2,\,3/2)$ basis that can be represented as a linear combination of Ir~$5d~t_{2g}$-orbitals and a mixture of up and down spin states. As seen in Fig.~\ref{fig:arpes}\,(c) and (e) DFT calculations already capture some of the low-energy spectral features. However, most noteworthy, they predict an Ir~$5d\,e_g$ electron pocket around $\Gamma$ and an Ir~$J_\text{eff}=3/2$ hole pocket around the A-point which are not seen in experiment. A way to reduce the discrepancy is to include short range Coulomb repulsion. Rather than looking for the lowest-energy solution within DFT+$U$, which would be a magnetic insulator at variance with experiment, we discard magnetism for this 9~uc film. In this framework, $U$ mainly acts to shift orbitals with different occupations relative to each other. We chose its value to match the position of the ARPES bands, thereby pushing the bands above (below) the chemical potential, whereupon the $e_g$ electrons are being predominantly transferred into the $J_\text{eff}=3/2$ band, leaving the $J_\text{eff}=1/2$ Luttinger volume relatively unchanged. It does not come as a surprise that the resulting values of $U$ and $J$ are significantly larger than \textit{ab initio} estimates using the constraint random phase approximation (cRPA) \cite{zhang_effective_2013, kim_dimensionality-strain_2017}. A more accurate treatment of the many-body processes based on dynamical mean field theory (DMFT) would most likely account for the experimental features with smaller values of the interaction (for DMFT studies for Sr$_2$IrO$_4$ see Refs. \onlinecite{arita_ab_2012, martins_reduced_2011}).\\%
Despite the overall good agreement a closer inspection of the theoretical DFT+$U$ band structure still reveals subtle remaining differences to the experimental data. In particular, the narrow band at the chemical potential between X- and M-point and the spectral weight near the R-point are not captured in the DFT+$U$ calculations in Fig.~\ref{fig:arpes}\,(d).  Indeed, previous ARPES studies using ultraviolet light have reported distinct narrow bands near the chemical potential resulting from backfolding due to octahedral rotations \cite{nie_interplay_2015}. As shown in Fig.~\ref{fig:OctRot}\,(a) collective rotations of IrO$_6$ octahedra introduce a periodic perturbation of the crystal potential, that enlarges the real-space unit cell by $2\times\sqrt{2}\times\sqrt{2}$ and correspondingly reduces the Brillouin zone as depicted in Fig.~\ref{fig:OctRot}\,(b). The resulting backfolding of bands is accompanied by an opening of small hybridization gaps at the Brillouin zone boundaries. In Fig.~\ref{fig:OctRot}\,(c) we present this band structure unfolded into the original Brillouin zone. The effect of the weak symmetry breaking due to octahedral rotations is taken into account by calculating the proper spectral weight distribution as described in Ref.~\onlinecite{ku_unfolding_2010} (represented by the size of the gray dots). The weighted, unfolded bands have a narrow bandwidth of $\approx 400\,\text{meV}$ and resolve the aforementioned discrepancies between experiment and theory by exhibiting Fermi crossings around the X- and R-point. Note that due to the breaking of translational symmetry the magnitude of octahedral rotations may be enhanced at the surface of thicker films, which could explain the more evident observation of a backfolded band structure in highly surface-sensitive ARPES measurements using He~I light \cite{nie_interplay_2015, liu_direct_2016} as opposed to bulk-sensitive photoemission in the soft x-ray regime \cite{yamasaki_three-dimensional_2016, DiracFeature}.\\%
%
The metastable, pseudo-cubic perovskite phase of SrIrO$_3$ can be regarded as the $n=\infty$ member of the Sr$_{n+1}$Ir$_n$O$_{3n+1}$ = ([SrIrO$_3$]$_n$,SrO) Ruddlesden-Popper series. These compounds essentially consist of $n$ SrIrO$_3$ perovskite layers, which are intercalated by SrO layers and laterally shifted with respect to each other such that no Ir-O-Ir bonds persist between neighboring [SrIrO$_3$]$_n$ blocks \cite{ruddlesden_compound_1958}. As shown above perovskite SrIrO$_3$ can be stabilized as epitaxial thin films, that essentially exhibit bulk electronic and structural properties \cite{longo_structure_1971,zhao_high-pressure_2008} above a thickness of at least 9 uc, i.e., paramagnetism and metallicity with a partially filled $J_\text{eff} = 1/2$ band. In contrast, the double- ($n=2$) and single-layer ($n=1$) compounds Sr$_3$Ir$_2$O$_7$ and Sr$_2$IrO$_4$ are insulators that exhibit a complex collinear and canted antiferromagnetic order, respectively. The intriguing variety of physical properties within this family of compounds is commonly explained by a dimensional crossover from the two-dimensional ($n=1$) to the three-dimensional limit ($n=\infty$). It is thus tempting to tune the dimensionality in SrIrO$_3$ thin films by reducing the number of monolayers and trace the electronic and structural changes.\\%
%
Figure~\ref{fig:MIT}\,(a) shows photoemission spectra (He\,I, $h\nu=21.2\,\text{eV}$) of SrIrO$_3$ films with thicknesses of $m=4,3,2,1$ and $0$ uc (bare Nb:SrTiO$_3$). Thicker films ($m\geq 4$) exhibit a metallic density of states with a pronounced Fermi-Dirac cut-off at the chemical potential as expected in the three-dimensional limit. Intriguingly, at $m=3$ the Fermi cut-off disappears and upon further reduction of the film thickness a distinct charge gap opens. Hence, in analogy to the Ruddlesden-Popper iridates the films undergo a metal-insulator transition as function of dimensionality as also observed in transport measurements \cite{groenendijk_spin-orbit_nodate}. As shown in the inset of Fig.~\ref{fig:MIT}\,(a) magnetic DFT+$U$ calculations for $m$ SrIrO$_3$ layers on 4 SrTiO$_3$ layers ($m$//4) similarly show a decreasing charge gap in the $k$ integrated density of states (DOS) as $m$ is increased. Note that in the presence of magnetic ordering the increasing film thickness alone does not trigger a transition from insulating to metallic in our calculations.\\
%
The photoemission gap opening is accompanied by a structural transition as inferred from low energy (LEED) and reflection high-energy electron diffraction (RHEED) from the film surface. As seen in Fig.~\ref{fig:MIT}\,(b) the surface periodicity changes from $\sqrt{2}\times\sqrt{2}$ ($m\leq 3$) to $2 \times 2$ ($m\geq 4$) with respect to the pseudo-tetragonal unit cell. Here, the doubling of the surface periodicity in the thick films simply reflects the complex rotational pattern of IrO$_6$ octahedra in the bulk as shown in Fig.~\ref{fig:MIT}\,(c), where the top layer of a SrIrO$_3$ film with $a^+b^-b^-$ bulk structure is depicted. However, for atomically thin films rotations about the [100] and [010] directions are suppressed since the SrIrO$_3$ film forms a corner-shared octahedral network with the SrTiO$_3$ substrate, which is a cubic perovskite without octahedral rotations at room temperature ($a^0a^0a^0$). Thus, only rotations about the surface normal ($a^0a^0b^{+/-}$ in Glazer notation) may prevail in thin films ($m\leq 3$), which result in a $\sqrt{2}\times \sqrt{2}$ surface periodicity. Note that the non-metallic Ruddlesden-Popper iridates Sr$_2$IrO$_4$ and Sr$_3$Ir$_2$O$_7$ also exhibit octahedral rotations exclusively about the $c$-axis reducing their space group symmetry from $I4/mmm$ to $I4_1/acd$ and $Bbcb$, respectively \cite{crawford_structural_1994, subramanian_single_1994, cao_anomalous_2002, kim_dimensionality_2012}.\\%
%
Figure~\ref{fig:Insulator}\,(a) shows the SX-ARPES bandmap of a 1 uc SrIrO$_3$ film grown on Nb:SrTiO$_3$ in comparison to the DFT+$U$ band structure of a $1//4$ SrIrO$_3$//SrTiO$_3$ slab along the pseudo-tetragonal high-symmetry line $\Gamma$-X-M-$\Gamma$. In excellent agreement with each other the experimental and theoretical data exhibit a flat band behavior with a valence band maximum at the M-point. Interestingly, as seen in the $k$-integrated DOS of the 1//4 slab in Fig.~\ref{fig:Insulator}\,(b), only the antiferromagnetic solution yields an insulating ground state, whereas the paramagnetic solution remains metallic like in the three-dimensional limit. This finding is in line with the enhanced spin fluctuations near the thickness-driven MIT recently observed in magnetoconductance measurements of samples identical to ours \cite{groenendijk_spin-orbit_nodate}. Similarly, the dimensionality-induced metal-insulator transitions observed in Ruddlesden-Popper iridate crystals \cite{jackeli_mott_2009,kim_dimensionality_2012,boseggia_antiferromagnetic_2012,boseggia_locking_2013, carter_theory_2013} and [(SrIrO$_3$)$_m$, SrTiO$_3$] superlattices \cite{matsuno_engineering_2015} are accompanied by a magnetic transition. Intriguingly, the antiferromagnetic DFT+$U$ band structure of the 1//4 SrIrO$_3$ slab shows a striking similarity to that of bulk Sr$_2$IrO$_4$ shown in Fig.~\ref{fig:Insulator}\,(c).\\%
%
For a deeper understanding of the driving mechanism behind the metal-insulator-transition one needs to take into account the subtle interplay between the dominant, comparably strong physical interactions ($U$, $W$, SOC) in $5d$ transition metal oxides, which leaves the electronic and magnetic ground state highly susceptible to small external perturbations. Here we have demonstrated that the SrIrO$_3$ film thickness can be used as experimental control parameter to tune three physical properties, which cooperatively determine the system's ground state. Firstly, the effective Coulomb interaction $U/W$ increases upon reduction of $m$ since the coordination of Ir sites becomes smaller, hence providing less hopping channels (smaller $W$) and less screening (bigger $U$) \cite{kim_dimensionality-strain_2017}. Secondly, the crystalline structure due to the IrO$_6$ rotations deviates from the rotational pattern of bulk SrIrO$_3$ in the two-dimensional limit, since the octahedral network with the cubic SrTiO$_3$ substrate imposes constraints upon the in-plane rotations. Finally, the magnetic ordering is susceptible to the dimensionality due to the strong competition between intra- and interlayer coupling \cite{kim_dimensionality_2012}.\\
The strong cooperative interplay between these degrees of freedom constitutes the complexity of the system. Specifically, octahedral rotations strongly affect the magnetic coupling in iridates due to pseudodipolar and Dzyaloshinsky-Moriya exchange interactions as evidenced by the locking of the Ir magnetic moments to the rotated oxygen octahedra in Sr$_2$IrO$_4$ \cite{boseggia_locking_2013}. In turn, the symmetry breaking due to octahedral rotations provides further spin-dependent hopping terms in the $J_\text{eff}$ basis that additionally increase the kinetic energy $W$ \cite{carter_theory_2013}. This tendency is reflected in the (albeit small) resistivity drop at $T=105\,\text{K}$ \cite{groenendijk_epitaxial_2016}, where the SrTiO$_3$ substrate undergoes a structural transition involving $a^0a^0c^-$-rotations of the TiO$_6$ octahedra \cite{shirane_lattice-dynamical_1969, glazer_classification_1972}, which can induce in-plane tiltings in the SrIrO$_3$ film depending on the domain structure in the SrTiO$_3$. On the other hand changes in $U/W$ will affect the magnetic ordering by altering the required critical on-site Coulomb repulsion $U$ for a magnetic transition to antiferromagnetic order \cite{carter_theory_2013}. The underlying reason for this extraordinary dimensionality-dependence is the spatially three-dimensional $J_\text{eff} = 1/2$ Kramers doublet wave function, which results from the mixing of orbitals of different symmetries with the spin degrees of freedom due to the strong spin-orbit coupling in $5d$ systems. This is in stark contrast to typical $3d$ systems like the cuprates, where the planar $e_g$ orbitals host the $S=1/2$ magnetic moments in the absence of strong spin-orbit coupling.\\
In this Letter we have illustrated how the electronic properties of materials with strong spin-orbit coupling and electronic correlations respond to the extraordinarily subtle interplay of collective octahedral rotations, magnetism, and dimensionality. We have demonstrated that thin film growth of such correlated quantum materials can be successfully employed to tune these parameters. With regard to Sr$_2$IrO$_4$ as a potential parent material for exotic superconductivity, the analogy between monolayer SrIrO$_3$ and bulk Sr$_2$IrO$_4$ may open a promising experimental avenue towards electron doping without the introduction of disorder through electrostatic or ion-liquid gating,  possibly pushing the system into a novel, spin-orbit driven superconducting phase.\\%
\\%
This work was supported by the Deutsche Forschungsgemeinschaft (SFB 1170 \textit{ToCoTronics} and FOR 1162). We thank D. Groenendijk, A. Caviglia (both at TU Delft), C. Autieri and S. Picozzi (both at CNR-SPIN, L'Aquila) for fruitful discussions as well as D. Pohl, A. Lubk, T. Gemming and B. B\"uchner (IFW Dresden) for transmission electron microscopy measurements. D.D.S. and G.S. gratefully acknowledge the Gauss Centre for Supercomputing e.V. (www.gauss-centre.eu) for funding this project by providing computing time on the GCS Supercomputer SuperMUC at Leibniz Supercomputing Centre (LRZ, www.lrz.de). M.C. acknowledges financial support from MIUR through the PRIN 2015 program (Prot. 2015C5SEJJ001) and SISSA/CNR project “Superconductivity, Ferroelectricity and Magnetism in bad metals”.
%


%

\end{document}